# Optical Limiting and Theoretical Modelling of Layered Transition Metal Dichalcogenide Nanosheets

Ningning Dong[1], Yuanxin Li[1], Yanyan Feng[1], Saifeng Zhang[1], Xiaoyan Zhang[1], Chunxia Chang[1], Jintai Fan[1], Long Zhang,[1] and Jun Wang[1,2,*]

[1]Key Laboratory of Materials for High-Power Laser, Shanghai Institute of Optics and Fine Mechanics, Chinese Academy of Sciences, Shanghai 201800, China.

[2]State Key Laboratory of High Field Laser Physics, Shanghai Institute of Optics and Fine Mechanics, Chinese Academy of Sciences, Shanghai 201800, China

*jwang@siom.ac.cn

Nonlinear optical property of transition metal dichalcogenide (TMDC) nanosheet dispersions, including $MoS_2$, $MoSe_2$, $WS_2$, and $WSe_2$, was performed by using Z-scan technique with ns pulsed laser at 1064 nm and 532 nm. The results demonstrate that the TMDC dispersions exhibit significant optical limiting response at 1064 nm due to nonlinear scattering, in contrast to the combined effect of both saturable absorption and nonlinear scattering at 532 nm. Selenium compounds show better optical limiting performance than that of the sulfides in the near infrared. A liquid dispersion system based theoretical modelling is proposed to estimate the number density of the nanosheet dispersions, the relationship between incident laser fluence and the size of the laser generated micro-bubbles, and hence the Mie scattering-induced broadband optical limiting behavior in the TMDC dispersions.

**Keywords:** Optical limiting; 2D transition metal dichalcogenide; number density in dispersions; Z-scan; Mie theory

**Introduction**

Under the promotion of research into graphene, two-dimensional (2D) nanomaterials have become one of the most widely studied fields in nanoscience.[1-5] Layered transition metal dichalcogenide (TMDC), as analogues of graphene, have attracted tremendous attention and been considered as potential candidate materials for photonic and optoelectronic devices owing to their extraordinary properties, such as ultrafast carrier dynamics, photoluminescence and electroluminescence, ultrafast nonlinear absorption, second and third harmonic generations, as well as indirect-to-direct band gap transition as bulk TMDC decreasing to monolayers.[4,6-13] Ultrafast nonlinear optical (NLO) property investigation is a fundamental but important aspect for the development of photonic and optoelectronic devices. For the purpose of developing diverse high performance photonic devices, it is actually essential to have a comprehensive understanding on the NLO properties of the potential working substances.

Recently, we reported the prominent broadband saturable absorption (SA) performance in layered TMDC nanosheets for fs and ps pulses over a broad wavelength range (ref. 14 and 15). Zhou et al. revealed the size-dependent NLO properties of thin $MoS_2$, $WS_2$, and $NbSe_2$ nanosheets for ps pulses at 532 nm (ref. 16). Fu et al. reported nonlinear SA in vertically stood $WS_2$ nanoplates using ps pulses at 532 nm (ref. 17). Wang et al. observed wavelength selective optical limiting effect in $MoS_2$ dispersions for fs pulses (ref. 18). TMDC-based mode-locking and Q-switching operations have been successfully demonstrated in a range of ultrashort pulsed laser systems.[14,19-22] As the most conventional laser source, ns pulses have been widely used in many science and technology fields. Thus, it is significant to understand the nonlinear interaction between intense ns pulses and TMDC nanosheets. So far, very few experimental studies on the NLO property of layered TMDC in the ns regime have been reported.

In this work, we prepared a series of layered TMDC nanosheets, including $MoS_2$, $MoSe_2$, $WS_2$, and $WSe_2$, in N-methyl-2-pyrrolidone (NMP) by using liquid-phase exfoliation technique. For comparison, graphene dispersions were prepared at the same time. Transmission electron microscopy (TEM), absorption spectroscopy and Raman spectroscopy were performed to characterize the quality of the layered nanostructures. We systematically investigated the NLO response of these 2D nanomaterials under the excitation of ns pulses at 1064 and 532 nm through a Z-scan apparatus. The TMDC dispersions exhibit significant nonlinear scattering induced optical limiting response at 1064 nm and 532 nm. Selenide compounds show better limiting performance than that of the sulfides in the near infrared. We propose a liquid dispersion system based theoretical modelling to estimate the number density in the nanosheet dispersions, the relationship between incident laser fluence and the size of the laser generated microbubbles, and hence the Mie scattering-induced broadband optical limiting behavior in the TMDC dispersions.

**Methods**

**Materials.** All chemicals used in this work were of analytical grade and used as supplied. Graphite flakes (product number 332461), $MoS_2$ (product number 234842) and $WS_2$ (product number 243639) powders, N-methyl-2-pyrrolidone (product number 328634) were purchased from Sigma-Aldrich. $MoSe_2$ (product number 778087) and $WSe_2$ (product number 13084) powders were purchased from Alfa Aesar and Sterm Chemicals, respectively.

**Preparation of nanosheet dispersions in NMP.** It has been proven that liquid-phase exfoliation is a simple and effective method to exfoliate bulk layered materials into mono- and/or few-layer 2D nanosheets with the help of appropriate dispersants.[7,23,24] In this work, high quality TMDC, including $MoS_2$, $MoSe_2$, $WS_2$, and $WSe_2$, as well as graphene dispersions in NMP were prepared with the similar procedure as our previous works.[25,26] The commercial TMDC powders were added in NMP with initial concentrations of 5 mg/ml, respectively. Ultra sonication was carried out through a point probe (flat head sonic tip) for 60 min with a power output of 285 W. To maintain sonication efficiency and prevent overheating, the samples were kept in ice-water bath. The resultant dispersions were centrifuged at 3000 rpm for 90 min to remove large aggregates, and the top two-thirds of the dispersions were gently extracted by pipetting. The obtained TMDC dispersions were stable against sedimentation over several weeks.

**Characterization.** The quality of the obtained TMDC nanosheets was characterized by TEM (Tecnai G2 F20 S-TWIN, FEI). UV-visible absorption spectra of the nanosheet dispersions in NMP were conducted using a PerkinElmer Lambda 750 instrument.

Raman spectroscopy measurements for the nanosheets (dried on $SiO_2$/Si wafers) were carried out using a Monovista-P optical workstation (a confocal microscopy system) with a LD pumped laser at 532 nm.

**Nonlinear optical measurements.** The NLO property of the TMDC dispersions were measured by using an open aperture Z-scan apparatus, which is widely adopted to investigate the nonlinear absorption, scattering and refraction processes.[27] The optical arrangement is similar to what we used in our previous works.[26,28] All experiments were performed using 6 ns pulses from a Q-switched Nd:YAG laser operating at 1064 nm and its second harmonic 532 nm, with the pulse repetition rate of 2 Hz. The laser beam was tightly focused with a 15 cm focus lens, and all dispersions were tested in 10×10 mm quartz cuvettes. Meanwhile, another focusing lens was setup at ~45° to the incident beam to collect the scattering signal from the dispersions. Three high-precision photo-detectors were used to monitor the reference, transmitted and scattering light, respectively. Focusing on practical applications, NLO samples should keep certain transmittance under low ambient light. In order to evaluate the NLO responses, these nanosheet dispersions were adjusted to have a same moderate linear transmittance ~80% at 1064 nm. At 532 nm, the corresponding linear transmittances were then ~63.8%, ~25.3%, ~55.1%, ~37.2% and ~79.7%, for the $MoS_2$, $MoSe_2$, $WS_2$, $WSe_2$ and graphene dispersions, respectively. The beam waist radii at the focus were estimated to be ~61 μm at 1064 nm and ~33 μm at 532 nm.[25]

**Results**

TEM was performed to analyze the status of these dispersed nanoflakes. Figures 1(a)-

(e) show the typical TEM images of MoS$_2$, MoSe$_2$, WS$_2$, WSe$_2$, and graphene nanoflakes. The sizes of the TMDC nanosheets are mostly below 500 nm which are significantly smaller than the graphene flakes with the average size of a few micrometers. Large quantities of TMDC flakes were observed as few-layer layered nanosheets, and aggregated particles are absent in these TEM images, confirming the high quality of the prepared liquid-phase exfoliated samples. Furthermore, monolayer or few-layer structures can be seen at the edge of the nanosheets, and then high-resolution TEM images were captured from these edge regions and followed by digital periodic filter processing (Figs. 1(f)-(j)). It appears that both TMDC and graphene possess hexagonally symmetric structures.

Figure 2(a) is the UV-visible absorption spectra of the TMDC and graphene dispersions. As expected, two typical characteristic absorption peaks of MoS$_2$, MoSe$_2$, WS$_2$, WSe$_2$ are clearly observed at the region of 500-900 nm, which correspond to the A1 and B1 direct excitonic transitions of the TMDC originated from the energy split of valence-band and spin-orbital coupling.[6,7,14,29-32] These two peaks indicate that the TMDC are dispersed in NMP as the 2H-phase. No prominent absorption peaks are observed for graphene. The Raman spectra of the TMDC and graphene nanosheets were performed in a Monovista-P optical workstation using a 532 nm excitation laser. The 521 cm$^{-1}$ phonon mode from the Si substrate was used for calibration. As seen in Fig. 2(b), the expected A$_{1g}$ mode and E$^1_{2g}$ mode are observed to be 406.7 cm$^{-1}$ and 381 cm$^{-1}$ in MoS$_2$, 241.8 cm$^{-1}$ and 287.5 cm$^{-1}$ in MoSe$_2$, 420 cm$^{-1}$ and 351.2 cm$^{-1}$ in WS$_2$ flakes, respectively.[12,24,33,34] For WSe$_2$, agreeing with the works of Tonndorf et al[12] and Terrones et al[13], we only find a single broad peak around 252 cm$^{-1}$ as a result of $A_{1g}$ mode and $E^1_{2g}$ mode degeneration in few layer WSe$_2$ nanosheets. At the same time, they observed that the peak position changes with layer number.[12, 13] Therefore, we deduce the widened peak in our experiment mainly originates from the conjunct effects of nanoflakes with different thicknesses prepared through liquid exfoliation technique.[23] Graphene is characterized with the D, G, and 2D peaks located at 1345 cm$^{-1}$, 1575 cm$^{-1}$, and 2700 cm$^{-1}$, respectively.[23,35-37] These results demonstrate that the prepared nanosheets are of high quality.

Figure 3 shows the excitation pulse energy dependent open-aperture Z-scan results of the TMDC and graphene dispersions at 1064 nm (top row) and 532 nm (bottom row), respectively. At 1064 nm, both TMDC and graphene dispersions exhibit a reduction in the transmittance on the focus of the lens, indicating a typical optical limiting property (Figs. 3(a)-(e)). In the TMDC, the value of the valley transmittance at the beam focus decreases gradually as the input pulse energy increasing. However, the value shows a slight rising after decreasing with the incident pulse energy clamping in graphene dispersions. This is preliminary considered to be associate with the low graphene nanosheet concentration in the dispersions. These prominent optical limiting behavior mainly originates from thermally induced nonlinear scattering.[26,28] Strong light scattering was observed when the dispersions passed through the focus of the incident beam (see Fig. 4). When 532 nm laser pulses were used, these samples displayed different NLO responses. Take MoS$_2$ for example, the normalized transmission curve shows a symmetrical peak with respect to the focus (z=0) at the lowest excitation energy of 10 µJ, indicating a SA mechanism in this sample. As the incident pulse energy increases, a valley inside the peak appears at the focus and becomes deeper gradually, which contributes to the nonlinear scattering (NLS) behavior occurring following SA in $MoS_2$. It should be pointed out that, at the highest excitation energy of 1500 µJ, only the valley can be observed and the peak disappears, which means NLS occupies a predominant role at the higher excitation pulse energy. The other three TMDC, i.e., $MoSe_2$, $WS_2$, and $WSe_2$, exhibit the similar phenomena as $MoS_2$ (Figs. 3(f)-(i)). Although we have reported the SA behaviors of graphene at femtosecond,[35,36] it is quite easy to generate NLS at nanosecond for graphene dispersions, and the SA performance can be buried by strong NLS at higher laser pulse excitation.[25,26] This is the reason that graphene only shows NLS induced optical limiting under the similar excitation condition (Fig. 3(j)).

Figure 4(a) depicts the nonlinear transmission and scattering as functions of incident fluence of the TMDC and graphene dispersions at 1064 nm. At the same level of linear transmittance (~80%) at 1064 nm, graphene possesses prominent optical limiting behavior at lower pulse energy excitation. However, TMDC show better optical limiting responses at higher pulse energy excitation, as shown in Figs. 3(a)-(e). Among the four TMDC, optical limiting responses follow the order $MoSe_2$ > $WSe_2$ > $MoS_2$ ~ $WS_2$. As shown in Fig. 4, the scattering signals increase significantly along with decrease of transmittance which implies that NLS is dominating the optical limiting performance in these 2D nanosheet dispersions. Similar optical limiting behavior at 532 nm (with different linear transmittances) are concluded, as shown in Fig. 4(b). It should be pointed out that the minimal transmittance ($T_{min}$) of TMDC decrease to ~0.2-0.3 at both 532 nm and 1064 nm, indicating their excellent attenuation effects over a broadband range from the visible to the near infrared. The optical limiting onset values ($F_{on}$, defined as the incident fluence at which optical limiting activity starts) and optical limiting threshold values ($F_{th}$, defined as the incident fluence at which the transmittance falls to 50% of the linear transmittance) for these samples at 1064 and 532 nm are summarized in Table 1. Among the five nanosheet dispersions, graphene possesses the minimum $F_{on}$ value which may related to the large nanosheet size and huge thermal conductivity ~$5.3 \times 10^3$ W/mK in graphene, which is about dozens of times larger in comparison with single layer $MoS_2$ (~103 W/mK), $MoSe_2$ (~54 W/mK), $WS_2$ (~142 W/mK) and $WSe_2$ (~53 W/mK) at room temperature.[38,39] Whereas TMDC possess quite lower $F_{th}$ values and $T_{min}$ values, these advantages make them to be potential candidates for broadband optical limiters at both the near infrared and the visible ranges.

**Discussion**

As mentioned above, the optical limiting response in these 2D nanomaterials at nanosecond is mainly attributed to NLS, and the scattering efficiency is largely dependent on the scattering cross section, hence relating to the size of scattering centers, i.e., micro-bubbles and/or micro-plasmas.[40] In our previous work, we have simulated the normalized transmission as a function of the radius of micro-bubbles by assuming different densities of graphene nanoflakes in dispersions.[25] As a consequence, it is vital to get the number density of the nanostructures for the investigation of NLS property. Here, we afford an effective method to estimate the nanosheet number density combing with the linear transmission. We treat the prepared dispersions as homogeneous media, and the nanosheets are uniform with the area of *S*, layer number of *n* in each suspension although the size and layer number show actually somewhat distributions in certain ranges. Ignoring the reflection and scattering, the transmittance $T_{nano}$ of an individual nanosheet can be expressed as $T_{nano} = e^{-\alpha nd}$, where $\alpha$ is the absorption coefficient of the nanosheet and *d* is the thickness of monolayer. And then, the absorbance of this nanosheet can be defined as $A' = 1 - T_{nano}$. Dividing the cuvette into *m* equal pieces along the laser transmitting direction (z axis), the thickness of each part can be expressed as $l = L/m$, *L* is the thickness of the quartz cuvette. If *m* is large enough, the part covered by the Gaussian beam in the dispersions can be considered as an area constituted by many pieces of circular slabs, and the thickness of each slab corresponds to *l*. Take one individual slab at $z_i$ (*i* =1, 2, …, m.) for example, the volume can be defined as $V_i = \pi R^2(z_i) \cdot l$, where $R(z_i)$ is the radius of this slab. The Gaussian beam can be seen as a normal distribution with the standard deviation $\sigma = \frac{1}{2}\omega(z_i)$, and $\omega(z_i)$ is the beam waist at $z_i$ with the expression of $\omega(z_i) = \omega_0 \cdot \sqrt{1 + (\frac{z_i \cdot \lambda}{\pi \omega_0^2})^2}$, $\lambda$ is laser wavelength, and $\omega_0$ is the beam waist radius of the Gaussian beam. Figure 5 gives a schematic about the process. Take $R(z_i) = 3\sigma$ (3$\sigma$ principle, the proportion of photons distributed in the circular equals to 99.73%), the number of the nanosheets in this slab can be given by

$$J = N_{eff} \cdot V_i = \frac{9}{4} \cdot (N_{eff} \pi l) \cdot \omega^2(z_i) \quad (1)$$

$N_{eff}$ is the effective nanosheet number per unit volume, say, effective nanosheet number density in the dispersions.

The Gaussian beam at the front face of this slab can be written as

$$I_{z_i}(x,y) = I_0 \cdot \left(\frac{\omega_0}{\omega_{z_i}}\right)^2 \cdot \exp[-\frac{2(x^2+y^2)}{\omega_{z_i}^2}] \quad (2)$$

and then the photon distribution should be

$$P_{z_i}(x,y) = \frac{I_{z_i}(x,y) \cdot \tau_p \cdot \pi R^2(z_i)}{h\nu} \quad (3)$$

the total photon number at the front face should be

$$P_{z_i,0} = 0.9973 \iint_{-\infty}^{+\infty} P_{z_i}(x,y)\,dxdy \quad (4)$$

Therefore, the total photon number at the front face of an individual nanosheet located at ($x_j$, $y_j$) in this slab will be [see Fig. 5(b)]

$$P_{z_i,nano} = \oiint_S P_{z_i}(x,y)ds = \frac{I_0 \cdot \tau_p \cdot \pi R^2(z_i)}{h\nu} \cdot \left(\frac{\omega_0}{\omega_{z_i}}\right)^2 \cdot S \cdot \exp[-\frac{2(x^2+y^2)}{\omega_{z_i}^2}] \quad (5)$$

The photon number absorbed by the individual nanosheet is

$$P_{z_i,j} = P_{z_i,nano} \cdot A' \quad (6)$$

The absorbed photon number in the whole slab is

$$P_{z_i,abs} = \sum_{j=1}^{J} P_{z_i,j} \quad (7)$$

The absorbance of this slab can be

$$A_i = \frac{P_{z_i,abs}}{P_{z_i,j}} = \frac{2S \cdot A'}{0.9973 \cdot \pi \omega_{z_i}^2} \cdot \sum_{j=1}^{J} \exp[-\frac{2(x^2+y^2)}{\omega_{z_i}^2}] \quad (8)$$

The transmittance of this slab is $T_i = 1 - A_i$. And then, the linear transmittance of the sample is

$$T_0 = \prod_{i=1}^{m} T_i = \prod_{i=1}^{m}(1 - A_i) \quad (9)$$

If we have the size and layer thickness information of these nanosheets which can be obtained through TEM and/or AFM characterizations, we can get the effective number density through the above analysis.

In addition, since the nanosheets are randomly oriented, the real nanosheet number density $N$ is not equals to the effective number density $N_{eff}$. Their relation can be deduced through the effective area of these nanosheets transmitted by the laser beam. As seen in Fig. 5(c), z is the laser transmitting direction, and a circular nanosheet with the radius $r$ locates in the o-xyz coordinate system with angles of $\theta$ and $\varphi$. The projection of the circular nanosheet in the xy plane is a ellipse with the radii of $r_x$ and $r_y$

$$r_x = r \cos \theta \tag{10}$$

$$r_y = r \sin \theta \cos \varphi \tag{11}$$

and the area of the ellipse is

$$S' = \pi r_x r_y = \frac{1}{2}\pi r^2 \sin 2\theta \cos \varphi \tag{12}$$

And then the ratio of the projected areas in the xy plane and their real areas of the nanosheets per unit volume can be written as

$$k = \frac{\frac{1}{2}\pi r^2 \cdot \sum_{n=1}^{N} \sin 2\theta_n \cos \varphi_n}{N \pi r^2} \tag{13}$$

By calculating, we find $k$ is ~0.2, and $N \approx 5 N_{eff}$.

In the above, we have mentioned that the sizes of the TMDC nanosheets are mostly less than 500 nm and graphene flakes are around several micrometers. Therefore, we suppose the areas of the TMDC and graphene nanosheets are 0.04 μm² and 1 μm², respectively, and the layer number equals to 5. Based on the assumptions, we can estimate the nanosheet number density of these dispersions. As we can see in Table 2, the calculated number densities at 1064 and 532 nm are not completely the same, but their ratio is as small as ~3 for the five kinds of dispersions. These acceptable difference implies the correctness and effectiveness of the proposed model. Furthermore, we can obtain the linear transmittance as a function of the nanosheet number density, area or layer number. As an example, we depict the simulation results of the MoS$_2$ nanosheet dispersions at 1064 and 532 nm in Fig. 6. At a certain linear transmittance, the larger the nanosheet area or the layer number is, the smaller number density required. With the model, one can estimate the nanosheet number density in a nanosheet dispersion from its linear transmittance and the nanosheet area/layer number.

Following the Beer-Lambert law, the decreased transmittance T$_{NL}$ of these dispersions can be expressed in the form of

$$T_{NL} = \exp(-\delta_{NL}NL) \qquad (14)$$

where $\delta_{NL}$ is the nonlinear extinction cross section. According to the theoretical simulation,[32] the scattering cross section increases significantly with the increasing size of micro-bubbles, meanwhile the absorption cross section decreases until it is negligible when the bubbles grow, effectively limiting the incident power. Therefore, we consider the micro-bubbles as non-absorbing dielectric spheres and the corresponding scattering cross section can be expressed by Mie theory as[41]

$$\delta = \frac{2\pi r'^2}{q^2}\Sigma_{l'=1}^{\infty}(2l' + 1)(|a_{l'}|^2 + |b_{l'}|^2) \qquad (15)$$

where $a_{l'}$ and $b_{l'}$ are the coefficients defined with Bessel function and its differentiation, $l'$ is an integer, $q$ is the corresponding size parameters, $r'$ is the radius of the micro-scatters. Substituting Equation 15 into Equation 14 allows one to estimate $T_{NL}$ as a function of the radius of micro-bubbles based on the calculated nanosheet number density. Figure 7 depicts the normalized transmittance variation with the micro-bubble radius as well as with the input laser fluence of these dispersions at 1064 nm and 532 nm. As we can see, the two curves fit very well, from which we can obtain the bubble size at certain laser fluence. That is to say, there is a correspondence between the bubble sizes and laser fluencies at certain dispersion. Figure 8 gives the bubble radius variation with the laser fluence for these dispersions, and the bubble radius increases gradually with the input laser fluence increasing. At a very low input laser fluence <1 J/cm$^2$, the bubbles in graphene dispersion have become large and the dispersion begin to emerge OL response. That is why graphene possesses superior optical limiting responses at lower excitation fluence. Whereas TMDC own large advantages at higher excitation fluence mainly due to their huge concentration although the micro-bubble sizes are quite small in comparison with graphene.

In summary, we have investigated the NLO behavior of MoS$_2$, MoSe$_2$, WS$_2$, WSe$_2$ and graphene nanosheets at nanosecond laser pulses. Both TMDC and graphene exhibit strong OL responses originated from NLS under the excitation of near infrared laser, while TMDC show a joint effect of SA and NLS at 532 nm. The fact that TMDC possess better OL responses at higher pulse energies than graphene make these 2D nanomaterials to be promising candidates for broadband optical limiters. In addition, we promote a theory analysis of the nanosheets number density dependent transmittance, and the relationship between incident laser fluence and micro-bubbles radius, which is helpful for the understanding of the NLS process in the nanosheet dispersions.

1. Geim, A. K. & Novoselov, K. S. The rise of graphene. *Nat. Mater.* **6**, 183–191 (2007).

2. Bonaccorso, F., Sun, Z., Hasan, T. & Ferrari, A. C. Graphene photonics and optoelectronics. *Nat. Photonics* **4**, 611 –622 (2010).

3. Mas-Balleste, R., Gomez-Navarro, C., Gomez-Herrero, J. & Zamora, F. 2D Materials: To graphene and beyond. *Nanoscale* **3**, 20–30 (2011).

4. Wang, Q. H., Kalantar-Zadeh, K., Kis, A., Coleman, J. N. & Strano, M. S. Electronics and optoelectronics of two dimensional transition metal dichalcogenides. *Nat. Nanotechnol.* **7**, 699–712 (2012).

5. Butler, S. Z. *et al.* Progress, challenges, and opportunities in two dimensional materials beyond graphene. *ACS Nano* **7**, 2898–2926 (2013).

6. Mak, K. F., Lee, C., Hone, J., Shan, J. & Heinz, T. F. Atomically thin $MoS_2$: a new direct-gap semiconductor. *Phys. Rev. Lett*. **105**, 136805 (2010).

7. Coleman, J. N. *et al.* Two-dimensional nanosheets produced by liquid exfoliation of layered materials. *Science* **331**, 568–571 (2011).

8. Ramakrishna Matte, H. S. S. *et al.* $MoS_2$ and $WS_2$ analogues of graphene. *Angew. Chem. Int. Ed.* **49**, 4059–4062 (2010).

9. Splendiani, A. *et al.* Emerging photoluminescence in monolayer $MoS_2$. *Nano Lett.* **10**, 1271-1275 (2010).

10. Zhao, W. J. *et al.* Origin of indirect optical transitions in few-layer $MoS_2$, $WS_2$, and $WSe_2$. *Nano Lett*. **13**, 5627-5634 (2013).

11. Zhang, Y. *et al.* Direct observation of the transition from indirect to direct bandgap in atomically thin epitaxial $MoSe_2$. *Nat. Nanotechnol.* **9**, 111-115 (2014).

12. Tonndorf, P. *et al.* Photoluminescence emission and Raman response of monolayer $MoS_2$,

MoSe$_2$, and WSe$_2$. *Opt. Express* **21**, 4908-4916 (2013).

13. Terrones, H. *et al.* New first order Raman-active modes in few layered transition metal dichalcogenides. *Sci. Rep.* **4**, 4215 (2014).

14. Wang, K. P. et al. Ultrafast saturable absorption of two-dimensional MoS$_2$ nanosheets. *ACS Nano* **7**, 9260-9267 (2013).

15. Wang, K. P. et al. Broadband ultrafast nonlinear absorption and nonlinear refraction of layered molybdenum dichalcogenide semiconductors. *Nanoscale* **6**, 10530-10535 (2014).

16. Zhou, K. G. *et al.* Size-dependent nonlinear optical properties of atomically thin transition metal dichalcogenide nanosheets. *Small* **11**, 694-701 (2015).

17. Fu, X. L., Qian, J. W., Qiao, X. F., Tan, P. H. & Peng, Z. J. Nonlinear saturable absorption of vertically stood WS$_2$ nanoplates. *Opt. Lett.* **39**, 6450-6453 (2014).

18. Wang, Y. Q., He, J., Xiao, S., Yang, N. A. & Chen, H. Z. Wavelength selective optical limiting effect on MoS$_2$ solution. *Acta. Phys. Sin.* **63**, 144204 (2014). Chinese.

19. Wang, S. X. Broadband few-layer MoS$_2$ saturable absorbers. *Adv. Mater.* **26**, 3538-3544 (2014).

20. Zhang, H. et al. Molybdenum disulfide (MoS$_2$) as a broadband saturable absorber for ultra-fast photonics. *Opt. Express* **22**, 7249-7260 (2014).

21. Liu, H. et al. Femtosecond pulse erbium-doped fiber laser by a few-layer MoS$_2$ saturable absorber. *Opt. Lett.* **39**, 4591-4594 (2014).

22. Wu, K., Zhang, X. Y., Wang, J. & Chen, J. P. 463 MHz fundamental mode-locked fiber laser based on few-layer MoS$_2$ saturable absorber. *Opt. Lett.* **40**, 1374-1377 (2015).

23. Hernandez, Y. *et al.* High-yield production of graphene by liquid-phase exfoliation of

graphite. *Nat. Nanotechnol.* **3**, 563–568 (2008).

24. Cunningham, G. *et al.* Solvent exfoliation of transition metal dichalcogenides: dispersibility of exfoliated nanosheets varies only weakly between compounds. *ACS Nano* **6**, 3468–3480 (2012).

25. Cheng, X. *et al.* Controllable broadband nonlinear optical response of graphene dispersions by tuning vacuum pressure. *Opt. Express* **21**, 16486-16493 (2013).

26. Wang, J., Hernandez, Y., Lotya, M., Coleman, J. N. & Blau, W. J. Broadband nonlinear optical response of graphene dispersions. *Adv. Mater.* **21**, 2430-2435 (2009).

27. Sheik-Bahae, M., Said, A. A., Wei, T. H., Hagan, D. J. & Stryland, E. W. V. Sensitive measurement of optical nonlinearities using a single beam. *IEEE J. Quantum Electron* **26**, 760-769 (1990).

28. Wang, J. & Blau, W. J. Solvent effect on optical limiting properties of single-walled carbon nanotube dispersions. *J. Phys. Chem. C* **112**, 2298-2303 (2008).

29. Liu, H. L. *et al.* Optical properties of monolayer transition metal dichalcogenides probed by spectroscopic ellipsometry. *Appl. Phys. Lett.* **105**, 201905 (2014).

30. Eda, G. & Maier, S. A. Two-dimensional crystals: managing light for optoelectronics. *ACS Nano* **7**, 5660-5665 (2013).

31. Bromley, R. A., Murray, R. B. & Yoffe, A. D. The band structures of some transition metal dichalcogenides: III. Group VIA: trigonal prism materials. *J. Phys. C: Solid State Phys.* **5**, 759-778 (1972).

32. Beal, A. R., Knights, J. C. & Liang, W. Y. Transmission spectra of some transition metal dichalcogenides: II. Group VIA: trigonal prismatic coordination. *J. Phys. C: Solid State*

*Phys.* **5**, 3540-3551 (1972).

33. Li, H. et al. From bulk to monolayer MoS$_2$: evolution of Raman scattering. *Adv. Funct. Mater.* **22**, 1385-1390 (2012).

34. Berkdemir, A. et al. Identification of individual and few layers of WS$_2$ using Raman spectroscopy. *Sci. Rep.* **3**, 1755 (2012).

35. Feng, Y. Y. et al. Saturable absorption behavior of free-standing graphene polymer composite films over broad wavelength and time ranges. *Opt. Express* **23**, 559-569 (2015).

36. Feng, Y. Y. et al. Host matrix effect on the near infrared saturation performance of graphene absorbers. *Opt. Mater. Express* **5**, 802-808 (2015).

37. Ferrari, A. C. *et al.* Raman spectrum of graphene and graphene layers. *Phys. Rev. Lett.* **97**, 187401 (2006).

38. Balandin, A. A. *et al.* Superior thermal conductivity of single-layer graphene. *Nano Lett.* **8**, 902-907 (2008).

39. Gu, X. K. & Yang, R. G. Phonon transport in single-layer transition metal dichalcogenides: a first-principles study. *Appl. Phys. Lett.* **105**, 131903 (2014).

40. Belousova, I. M., Mironova, N. G. & Yur'ev, M. S. Theoretical investigation of nonlinear limiting of laser radiation power by suspensions of carbon particles. *Opt. Spectrosc.* **94**, 86–91 (2003).

41. He, G. S., Qin, H. Y. & Zheng, Q. D. Rayleigh, mie, and tyndall scatterings of polystyrene microspheres in water: wavelength, size, and angle dependences. *J. Appl. Phys.* **105**, 023110 (2009).

**Acknowledgements**

This work is supported in part by NSFC (No. 61178007 and 61308087), the External Cooperation Program of BIC, CAS (No. 181231KYSB20130007), Science and Technology Commission of Shanghai Municipality (No. 12ZR1451800) and the Excellent Academic Leader of Shanghai (No. 10XD1404600). J.W. thanks the National 10000-Talent Program and CAS 100-Talent Program for financial support, and acknowledges Prof. Werner J. Blau at Trinity College Dublin for his helpful discussion in this work.

**Author contributions**

N.D. and J.W. designed research. N.D. prepared the samples. Y.F. and J.F. performed the TEM measurements. Y.L. contributed to the absorption spectra and Raman spectroscopy characterization, and designed the theoretical model. N.D. conducted the nonlinear optical experiments and analyzed the data. N.D. and J.W. wrote the paper. All authors discussed the results and the interpretation.

**Additional information**

Competing financial interests: The authors declare no competing financial interests.

**Figure 1**

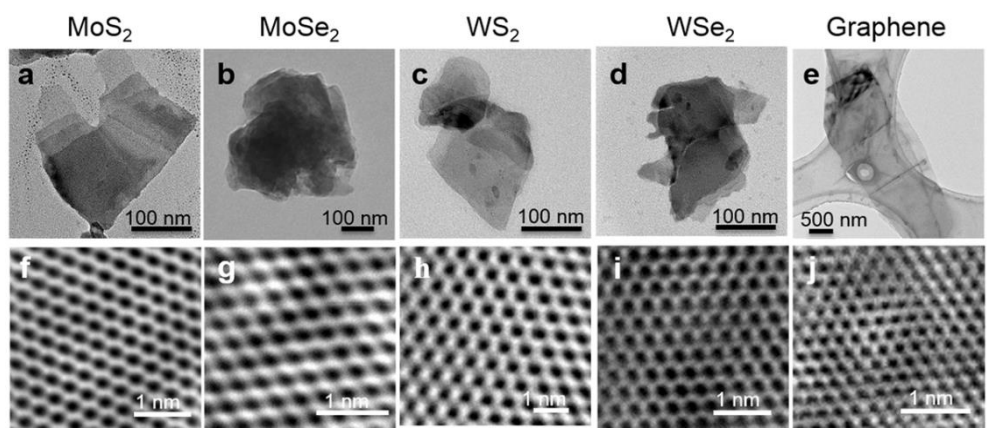

**Figure 1** Low-resolution TEM images (top row) and high-resolution TEM images (bottom row) of the $MoS_2$, $MoSe_2$, $WS_2$, $WSe_2$, and graphene nanoflakes.

**Figure 2**

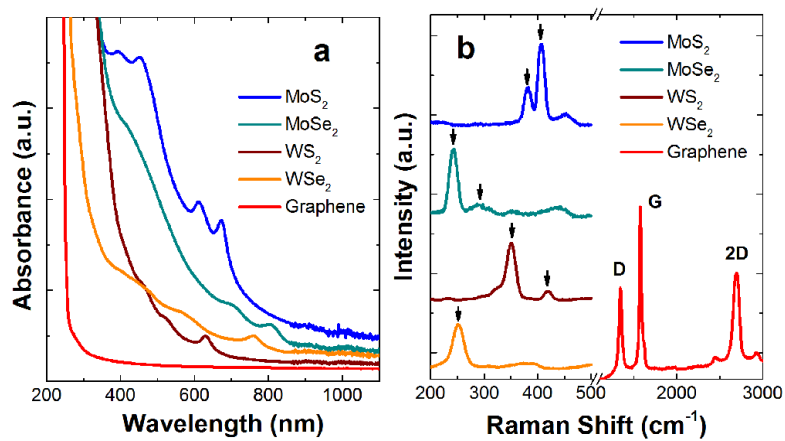

**Figure 2** (**a**) UV-visible spectra of the TMDC (MoS$_2$, MoSe$_2$, WS$_2$, and WSe$_2$) and graphene dispersions. (**b**) Raman spectra of the dried TMDC and graphene films.

**Figure 3**

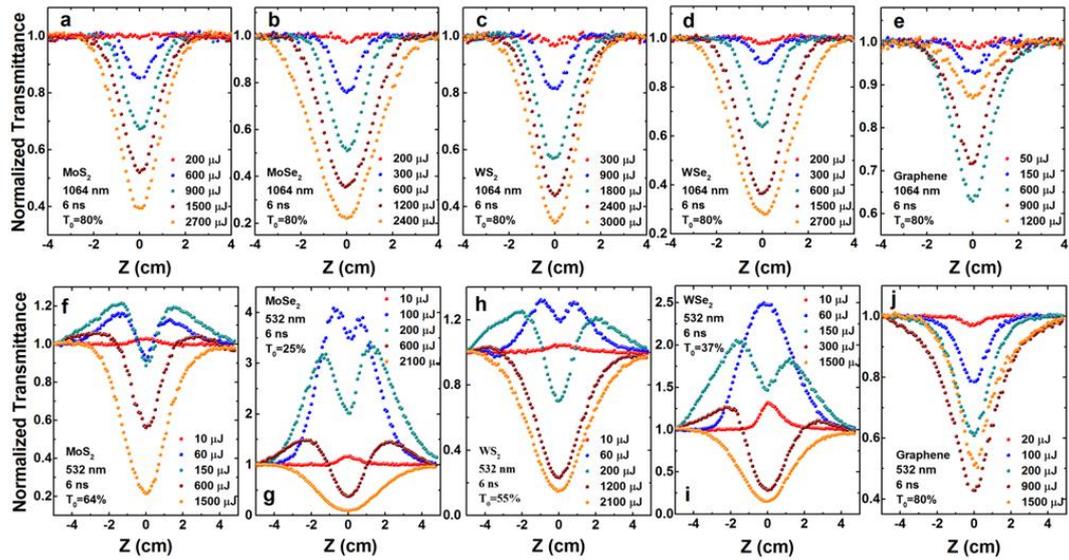

**Figure 3** Typical open-aperture Z-scan data with normalized transmittance as a function of the sample position Z for the $MoS_2$ (**a**, **f**), $MoSe_2$ (**b**, **g**), $WS_2$ (**c**, **h**), $WSe_2$ (**d**, **i**), and graphene (**e**, **j**) dispersions in NMP at 1064 and 532 nm, respectively, with different incident laser pulse energy.

**Figure 4**

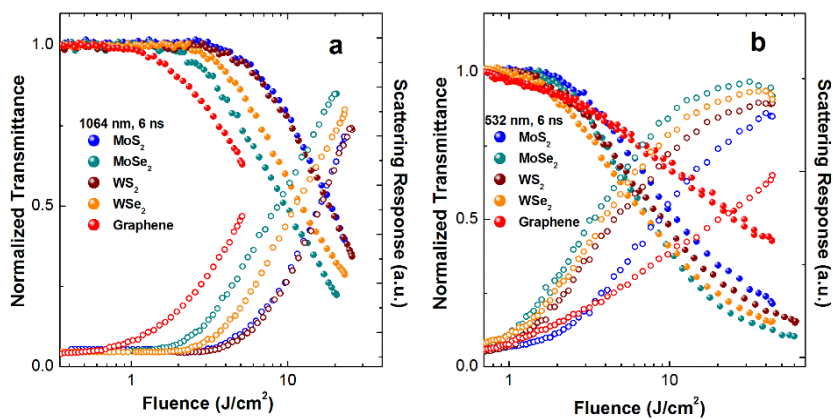

**Figure 4** Normalized transmittance (solid circles) and scattering response (open circles) of these nanosheet dispersions at (**a**) 1064 and (**b**) 532 nm, respectively.

**Figure 5**

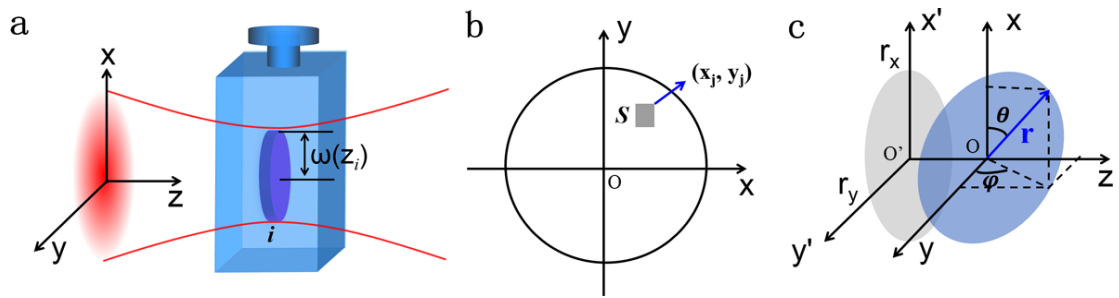

**Figure 5** (**a**) A piece of slab and (**b**) a nanosheet as a model to explain the calculation process. (**c**) The schematic to calculate the projected area at xy plane of a nanosheet placed with angles of $\theta$ and $\varphi$ in o-xyz coordinate system.

**Figure 6**

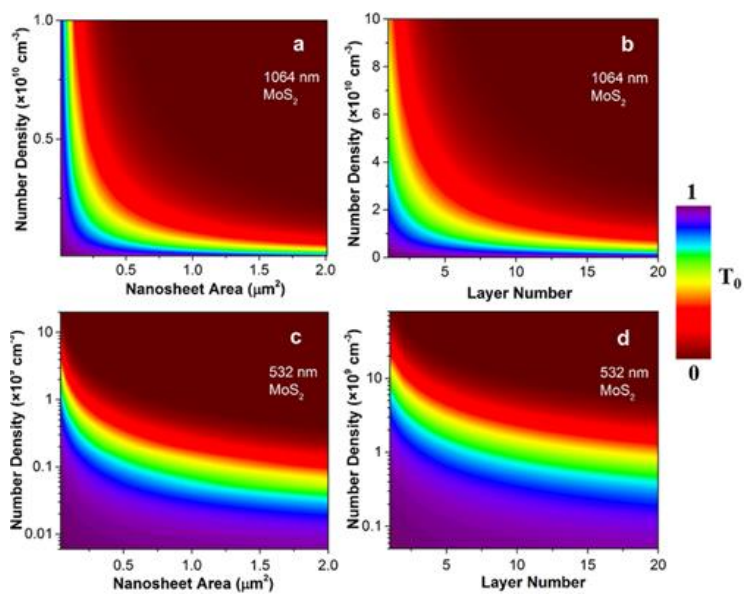

**Figure 6** Linear transmittance mapping of $MoS_2$ dispersions in $10\times10$ mm quartz cuvette at 1064 and 532 nm nanosecond laser pulses. (**a, c**) Variation in nanosheet number density and area with a constant layer number of 5. (**b, d**) Variation in nanosheet number density and layer number with a constant area of 0.04 $\mu m^2$. The color bar represents the linear transmittance.

**Figure 7**

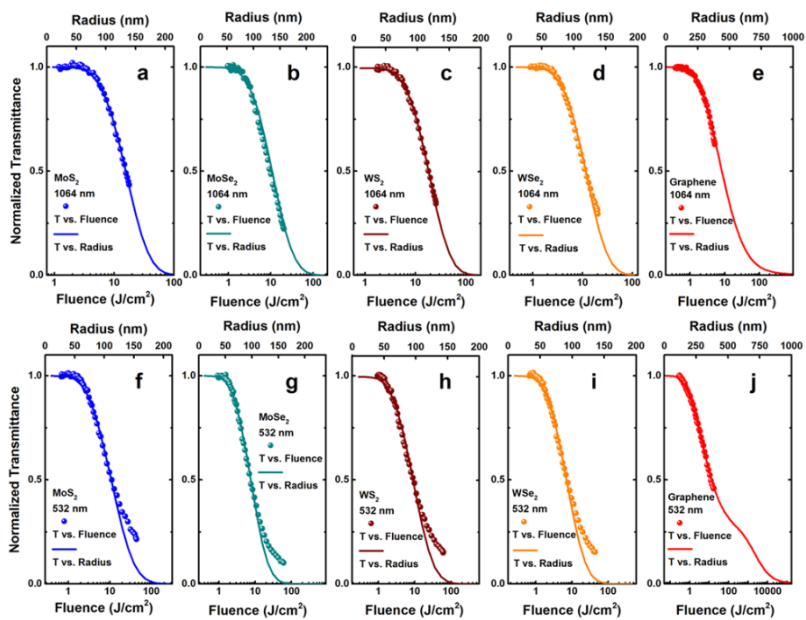

**Figure 7** Normalized transmittance as a function of input laser fluence (solid dots) and the radius of micro-bubbles (solid line) for the nanosheet dispersions at 1064 (top row) and 532 nm (bottom row).

**Figure 8**

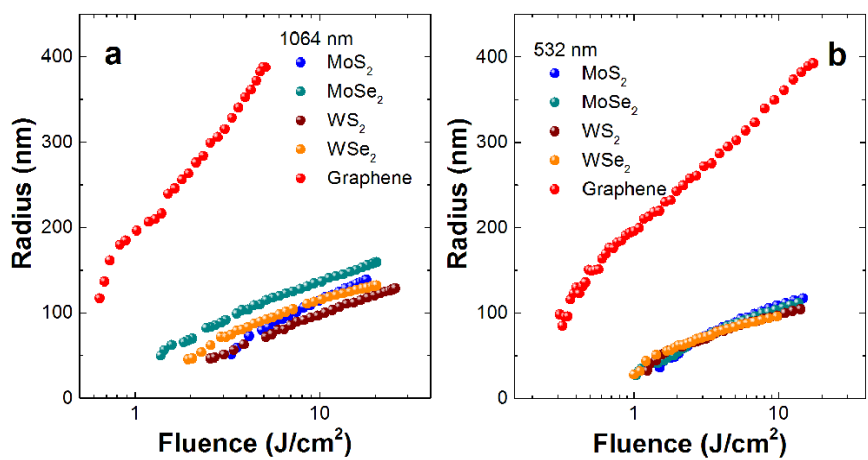

**Figure 8** The radius of micro-bubbles variation with the input laser fluence for the nanosheet dispersions at (**a**) 1064 and (**b**) 532 nm, respectively.

**Table 1** Linear transmittance ($T_0$), optical limiting onsets ($F_{on}$) and optical limiting thresholds ($F_{th}$) of these nanosheet dispersions at 1064 and 532 nm.

| Sample | 1064 nm | | | 532 nm | | |
|---|---|---|---|---|---|---|
| | $T_0$ | $F_{on}$ (J/cm$^2$) | $F_{th}$ (J/cm$^2$) | $T_0$ | $F_{on}$ (J/cm$^2$) | $F_{th}$ (J/cm$^2$) |
| MoS$_2$ | 79.7% | 3.28 | 19 | 63.8% | 1.52 | 11.16 |
| MoSe$_2$ | 79.8% | 1.37 | 9.8 | 25.3% | 1.47 | 7.3 |
| WS$_2$ | 79.6% | 2.56 | 18.25 | 55.1% | 1.24 | 9.35 |
| WSe$_2$ | 80.2% | 2.3 | 12 | 37.2% | 0.99 | 7.2 |
| Graphene | 80.1% | 0.64 | - | 79.7% | 0.44 | 15.15 |

**Table 2** The calculated nanosheet number density (cm$^{-3}$) at 1064 and 532 nm.

| Number density | MoS$_2$ | MoSe$_2$ | WS$_2$ | WSe$_2$ | Graphene |
|---|---|---|---|---|---|
| $N_{1064}$ | 1.17×10$^{10}$ | 1.09×10$^{10}$ | 2.15×10$^{10}$ | 2.07×10$^{10}$ | 8.30×10$^7$ |
| $N_{532}$ | 4.61×10$^9$ | 8.05×10$^9$ | 8.90×10$^9$ | 1.11×10$^{10}$ | 7.10×10$^7$ |